**Development of hard water sensor using Fluorescence Resonance Energy Transfer**

Dibyendu Dey, D. Bhattacharjee, S. Chakraborty, Syed Arshad Hussain*

Department of Physics, Tripura University, Suryamaninagar – 799022, Tripura, India

* Corresponding author

Email: sa_h153@hotmail.com, sah.phy@tripuraniv.in

Ph: +919862804849 (M), +91381 2375317 (O)

Fax: +913812374802

ABSTRACT

A method is presented for the sensing of water hardness by determining the concentration of calcium and magnesium in water, based on Fluorescence resonance energy transfer (FRET) process. The principle of the proposed sensor is based on the change of FRET efficiency between two laser dyes Acriflavine (Acf) and Rhodamine B (RhB) in presence of permanent hard water components ($CaCl_2$ and $MgCl_2$). Nanodimensional clay platelet laponite was used to enhance the efficiency of the sensor.

# 1. Introduction

"Hard water" has high mineral content in compare to "soft water". Generally the hard water is not harmful to one's health, but can cause serious problems in industrial settings, where water hardness should be monitored to avoid breakdowns of the costly equipments that handle water. The hardness of water is determined by the concentration of multivalent cations (positively charged metal complexes with a charge greater than 1+) in water. The most common cations found in hard water include $Ca^{2+}$ and $Mg^{2+}$. The presence of dissolved carbonate minerals ($CaCO_3$ and $MgCO_3$) provide a temporary hardness in water, which can be reduced either by boiling the water or by addition of lime (calcium hydroxide) [1]. On the other hand the dissolved chloride minerals ($CaCl_2$ and $MgCl_2$) cause the permanent hardness of water that can not be removed easily, because it becomes more soluble as the temperature increases [2]. In that sense it is very important to identify the permanent hardness of water before use. One of the most useful steps to water analysis is the determination of the concentration of calcium and magnesium ions, whether individually or overall hardness. In the routine laboratories the volumetric methods are the most commonly used methods for water analysis. Now a days the involvement of absorption or fluorescence spectroscopy for water analysis has received particular attention [3]. Sweetser and Bricker were the first who used the spectroscopic measurements to determine the concentration of calcium and magnesium ions in water [3]. Ion chromatography (IC) is another very powerful method for the analyses of anions and cations in aqueous solution [4, 5]. Argüello and Fritz reported a method for the separation of $Ca^{2+}$ and $Mg^{2+}$ in hard water samples based on ion-chromatography and spectroscopic method [6]. E. Gömez et al reported a method for the simultaneous spectroscopic determination of calcium and magnesium using a diode-array detector [7].

The Fluorescence resonance energy transfer (FRET) phenomenon may be very effective tool for the designing of hard water sensors. Based on the FRET between two laser dyes here we demonstrated a hard water sensor. To the best of our knowledge this could be the first attempt, where FRET process has been used for the detection of the hardness of water. FRET between two molecules is an important physical phenomenon, where transfer of energy from an excited flurophore to a suitable acceptor flurophore occurred [8, 9]. This technique is very important for the understanding of some biological systems and has potential applications in optoelectronic and thin film devices [10-14]. Combining FRET with optical microscopy, it is possible to determine the approach between two molecules within nanometers. The main requirements for the FRET to occur are (i) sufficient overlap between the absorption band of acceptor fluorophore and the fluorescence band of donor fluorophore and (ii) both the donor and acceptor molecule must be in close proximity of the order of 1–10 nm [8, 9]. The intervening of solvent or other macromolecules has little effect on the FRET efficiency. If the distance between the donor–acceptor changes then FRET efficiency also changes.

Here in the process of designing hard water sensor based on FRET process, we have used two dyes Acriflavine (Acf) and Rhodamine B (RhB) as energy donor and acceptor. In principle both the dyes are suitable for fluorescence resonance energy transfer. Both the dyes are highly fluorescent and the fluorescence spectrum of Acf sufficiently overlaps with the absorption spectrum of RhB. P. D. Sahare et al [15] reported the fluorescence resonance energy transfer in binary solution mixture using these two dyes. In one of our earlier work we have demonstrated a pH sensor based on the FRET between Acf and RhB [16]. The energy transfer efficiency has been effected if the distance between the donor – acceptor pair has been altered due to the presence of any external agency or change of the microenvironment. It has been observed that when distance between fluorophores (dyes) is decreased due to adsorption on to nanoclay sheet, the FRET efficiency increases [17, 18].

In the present communication we tried to investigate the effect of $Mg^{2+}$ or $Ca^{2+}$ or both on the FRET efficiency between two fluorophores, Acf and RhB in presence of nanoclay sheet laponite. Here we have chosen $Mg^{2+}$ or $Ca^{2+}$ because the presence of these two cations mainly determines the extent of hardness of the water. Our investigation showed that FRET efficiency decreases with increasing salt concentration. It has also been demonstrated that with proper calibration, FRET between Acf and RhB can be used to sense the hardness of water.

**2. Materials and methods**

2.1. Material

Both the dyes Acriflavine (Acf) and Rhodamine B (RhB) were purchased from Sigma Chemical Co., USA and used as received. Ultrapure Milli-Q water (resistivity 18.2 MΩ-cm) was used as solvent. The dyes used in our studies are cationic in nature. The clay mineral used in the present work was Laponite, obtained from Laponite Inorganic, UK and used as received. The size of the clay platelet is less than 0.05 μm and CEC is 0.739 meq/g determined with CsCl [19]. Both $MgCl_2$ and $CaCl_2$ were purched from Thermo Fisher Scientific India Pvt. Ltd. and used as received. Dye solutions were prepared in Milli-Q water. For spectroscopic measurement the solution concentration was optimized at $10^{-6}$ M. The clay dispersion was prepared by using Millipore water and stirred for 24 hours with a magnetic stirrer followed by 30 minutes ultrasonication before use. The concentration of clay was kept fixed at 2 ppm through out the experiment. To check the effect of clay on the spectral characteristics the dye solutions (Acf and RhB) were prepared in the clay suspensions (2 ppm). The dye concentration was maintained at $10^{-6}$ M. In order to check the effect of salt on spectral characteristics in presence of clay, first of all the salts were added in the clay dispersion at different concentration. Then the dyes were added in the salt mixed clay dispersions. In all cases the clay concentration was 2 ppm and the dye concentration was $10^{-6}$ M.

2.2. UV−Vis absorption and fluorescence spectra measurement

UV−Vis absorption and fluorescence spectra of the solutions were recorded by a PerkinElmer Lambda-25 Spectrophotometer and PerkinElmer LS-55 Fluorescence Spectrophotometer respectively. For fluorescence measurement the excitation wavelength was 420 nm (close to the absorption maxima of Acf).

**3. Results and discussions**

3.1. UV−Vis absorption and steady state fluorescence spectra study

The absorption and emission maxima of Acf are centered at 449 and 502 nm respectively which is assigned due to the Acf monomers [15]. On the other hand RhB absorption spectrum possess prominent intense 0-0 band at 553 nm along with a weak hump at 520 nm which is assigned due to the 0-1 vibronic transition [20]. The RhB fluorescence spectrum shows prominent band at 571 nm which is assigned due to the RhB monomeric emission [20]. The corresponding absorption and emission spectra of the above results are shown in figure 1 of the supporting information.

3.2. FRET between Acf and RhB in aqueous solution with salt

To study the energy transfer between Acf and RhB the fluorescence spectra of Acf and RhB mixture (1:1 volume ratio) were measured with excitation wavelength fixed at 420 nm (close to the absorption monomer of Acf). Figure 1 shows the fluorescence spectra of Acf, RhB and their mixture in water solution in presence and absence of salt. From the figure it was observed that the fluorescence intensity of pure Acf (curve 5, of figure 1) is much higher, on the other hand the fluorescence intensity of pure RhB (curve 6, of figure 1) is almost negligible. This is because the excitation wavelength (420 nm) was chosen in order to excite the Acf molecule directly and to avoid the direct excitation of the RhB molecule. However, the Acf-RhB mixture fluorescence spectrum is very interesting. Here the Acf emission decreases with respect to pure Acf and on the other side RhB emission increases with respect to pure RhB (curve 1, of figure 1). This is mainly due to the transfer of energy from Acf molecule to RhB molecule via fluorescence resonance energy transfer. In order to confirm this, excitation spectra was recorded with monitoring emission wavelength 500 nm (Acf emission maximum) and 571 nm (RhB emission maximum) and observed that both the excitation spectra are very similar to the absorption spectrum of Acf monomer (figure 2 of supporting information). This confirms that the RhB fluorescence is mainly due to the light absorption by Acf and corresponding transfer to RhB monomer. Thus FRET between Acf to RhB has been confirmed.

Our main purpose is to detect $CaCl_2$ and $MgCl_2$ or their mixture as a permanent hard water component in aqueous solution by using the FRET process between Acf and RhB. There are few studies where the effect of some chloride salts in FRET has been studied. Loyse M. Felber et al studied the effect of NaCl on FRET between CFP and YFP [21]. The effect of chloride ions and similar halide ions results a decrease in FRET efficiency at pH close to its pKa value [22, 23]. Ken-ichi Yoshioka et al studied the self assembly of HsRed51 by measuring the FRET from the fluorescein-labeled protein to the Rhodamine-labeled protein which is dependent strongly on high salt concentration indicating the ionic interaction between positively and negatively charged aminoacids [24].

Here we have studied the effect of $CaCl_2$ and $MgCl_2$ and their mixture on the transfer of energy from Acf to RhB in their aqueous solution. Fluorescence spectra of Acf-RhB mixture in presence of $CaCl_2$, $MgCl_2$ and their mixture ($CaCl_2+MgCl_2$) are also shown in figure 1 (curve 2, 3 and 4). For all the spectra shown in figure 1 the concentration of $CaCl_2$/$MgCl_2$/their mixture was kept at 0.06 mg/ml, which is the initializing concentration of hard water known as moderately hard water [1, 2]. From figure 1 it was observed that the transfer of energy from Acf to RhB was decreased in presence of salt (curve 2, 3 and 4). Based on the fluorescence spectra the fluorescence energy transfer efficiency have been calculated using the following equation [25]

$$E = 1 - \frac{F_{DA}}{F_D}$$

Where $F_{DA}$ is the relative fluorescence intensity of the donor in the presence of acceptor and $F_D$ is the fluorescence intensity of the donor in the absence of the acceptor.

It has been observed that the FRET efficiency in aqueous solution is 11.37% which decreases to 1.7% and 5.2% for the presence of $CaCl_2$ and $MgCl_2$ respectively. The aqueous solutions of the salts generate the cationic $Ca^{2+}$ and $Mg^{2+}$ ions. The laser dyes Acf and RhB both are cationic in nature and repeal each other in aqueous solution. The introduction of cationic $Ca^{2+}$ and $Mg^{2+}$ ions in the solution may cause an increase in the electrostatic repulsion between Acf and RhB molecules which can result in a large intermolecular separation. Accordingly, the FRET efficiency decreases. On the other hand the diameter of $Ca^{2+}$ ion is larger in compared to $Mg^{2+}$ ion which could be responsible for a small but noticeable variation in FRET between Acf and RhB in presence of $Ca^{2+}$ and $Mg^{2+}$ ions. It is also observed that with the increase in salt concentration the transfer of energy from Acf to RhB decreases further (figure not shown). In presence of both $Ca^{2+}$ and $Mg^{2+}$ ions, the FRET efficiency also decreases and the efficiency is 4.38% which lies in between the efficiencies for the presence of either $Ca^{2+}$ or $Mg^{2+}$.

3.3. FRET between Acf and RhB in clay dispersion with salt

Our previous investigations revealed that the energy transfer occurred from Acf to RhB in aqueous solution. Also in presence of salt ($Ca^{2+}$ or $Mg^{2+}$) the energy transfer efficiency decreases. However the energy transfer efficiency as well as the change in efficiency in presence of salt is very small due to the large intermolecular separation between Acf and RhB. In the present study our aim is to design a sensor which can sense the presence of $Ca^{2+}$ or $Mg^{2+}$ or both by observing the change in FRET efficiency. Accordingly it is very important to have large FRET efficiency between Acf and RhB as well as noticeable change in FRET efficiency between Acf and RhB due to the introduction of hard water components ($Ca^{2+}$ or $Mg^{2+}$), so that we can use it as a hard water sensor with minimum error level. Here in the designed sensor the hardness of the sample water will be sensed just by observing the change in the FRET efficiency. Accordingly in order to enhance the FRET efficiency we have incorporated nano clay laponite in Acf-RhB mixture. It has been observed that the FRET efficiency increases in presence of laponite particle.

It is important to mention in this context that in one of our earlier works it has been observed that the presence of nanoclay laponite increased the FRET efficiency between N,N'-dioctadecyl thiacyanine perchlorate (NK) and octadecyl rhodamine B chloride (RhB) [18, 26]. Effect of nanoclay laponite on the energy transfer efficiency between Acf and RhB has also been studied [16]. It has been observed that the presence of clay platelet increase the energy transfer efficiency. Fluoresecence spectra of Acf, RhB and their mixture in absence and presence of salt in aqueous clay dispersion are shown in figure 2. Here in presence of laponite the FRET efficiency has increased to 78.17% which was 11.37% in absence of clay. This has been evidenced from the observed decrease of Acf fluorescence in favour of RhB fluorescence intensity in presence of nanoclay platelets (Curve-1 of figure 2).

It is worthwhile to mention in this context that clay particles are negatively charged and have layered structure with a cation exchange capacity [27, 28]. Both the dyes Acf and RhB under investigation are positively charged. Accordingly they are adsorbed on to the clay layers [27, 28]. On the other hand FRET process is very sensitive to distances between the energy donor and acceptor and occurs only when the distance between the D-A pair is of the order of 1-10 nm [8, 9]. Therefore in the present case, clay particles play an important role in determining the concentration of the dyes on their surfaces or to make possible close interaction between energy donor and acceptor in contrast to the pure aqueous solution. Now our main purpose is to observe the change in FRET between Acf and RhB in clay dispersion due to the introduction of $CaCl_2$/$MgCl_2$/their mixture. From figure 2 (curve 2, 3 and 4) it was observed that the transfer of energy from Acf to RhB decreases quite remarkably due to the introduction of $CaCl_2$/$MgCl_2$/their mixture for the concentration of 0.06 mg/ml in presence of clay. It was observed that the transfer of energy is much smaller due to the presence of $CaCl_2$ in compare to $MgCl_2$. (Table 1 summarize the calculated efficiencies).

The decrease in FRET between Acf and RhB in presence of hard water components, must involve the reaction of the cations ($Ca^{2+}$ and $Mg^{2+}$) present in hard water with the clay minerals through cation exchange reaction. Marshall [29] formulated that the replacement of cations on a charged clay mineral surface by those present in a contact solution takes place according to the order of bonding energy of the common metal cations. In general, this bonding energy is of the order: Ca > Mg > K > H > Na. Thus the probability of adsorption of $Ca^{2+}$ in clay suspension is more than $Mg^{2+}$. The tendency of $Ca^{2+}$ ion to interact with the negatively charged clay layers is more compared to $Mg^{2+}$ ion of same concentration of both clay suspension and salt solution. Accordingly, most of the negative charges in the clay surfaces are neutralized by $Ca^{2+}$ ion compare to $Mg^{2+}$ ion and there exists very few unoccupied negative charges on the clay surface for the cationic dye molecules to be adsorbed. As a result the separation between the cationic dye molecules increases more in $CaCl_2$ solution rather than $MgCl_2$ leading to a less FRET in presence of $CaCl_2$ compare to $MgCl_2$.

3.4. Schematic diagram

A schematic diagram showing the organization at Acf and RhB in absence and presence of clay laponite and salt is shown in figure 3. Normally in absence of clay and salt the distance between Acf and RhB molecules in aqueous solution is larger resulting lower energy transfer efficiency (figure 3a). In presence of clay the dyes are adsorbed by cation exchange reaction on to the clay surface and accordingly the distance between Acf and RhB decreases (figure 3b). These results can increase the energy transfer efficiency. In presence of both clay and salt, the probability of adsorption of $Ca^{2+}$ and $Mg^{2+}$ cations are larger in compared to the cationic dyes (figure 3c). This is because in the process of dye-clay-salt solution preparation initially the salt was added to the clay dispersion followed by the dye addition. Accordingly, most of the negative charges on the clay surface are neutralized by the $Ca^{2+}$ and $Mg^{2+}$ cations and there exist very few unoccupied negative charges on the clay surface to adsorb the cationic dyes. Accordingly the average distance between Acf and RhB become larger. This results a decrease in energy transfer efficiency.

3.5. Effect of variation of salt concentration on FRET efficiency

In the previous sections we have seen that presence of laponite particle increases the FRET efficiency between Acf and RhB. Whereas, presence of $Mg^{2+}$ or $Ca^{2+}$ or both decreases the FRET efficiency. Now in order to check the effect of variation of salt concentration ( extent of hardness) on the FRET efficiency, we have measured the fluorescence spectra of Acf and RhB mixture with different salt ($MgCl_2$ + $CaCl_2$) concentration in presence of clay laponite and the FRET efficiency have been calculated.

Figure 4 shows the fluorescence spectra of Acf–RhB mixture in presence of clay with varying amount of salt concentration viz. 0.05 mg/ml (soft water), 0.06 mg/ml (moderately hard water) and 0.12 mg/ml (very hard water). From the figure it has been observed that the RhB fluorescence intensity decreases with increase in salt concentration. Also the corresponding Acf fluorescence intensity increases. This indicates that with increase in salt concentration FRET efficiency between Acf and RhB decreases. The plot of FRET efficiency as a function of salt ($MgCl_2$ + $CaCl_2$) concentration (inset of figure 4) clearly shows that FRET efficiency decreases with increasing salt concentration (ranging from 0.03 mg/ml to 0.2 mg/ml). The values of FRET efficiencies with salt concentration are listed in table 2. This result suggests that it is possible to sense the hardness of water by observing the change in FRET efficiency with salt concentration.

3.6. Design of hard water sensor

Based on the variation of FRET efficiency or fluorescence intensity, depending on the salt concentration we have demonstrated a hard water sensor. In the process of hard water sensing first of all clay (laponite) dispersion will be prepared using the sample water followed by addition of dyes (Acf and RhB). Then the fluorescence spectra of the solution will be measured. By observing the fluorescence intensity or FRET efficiency calculated from the observed fluorescence spectra it would be possible to sense the hardness of the sample water.

Figure 5a and b show the plot of FRET efficiency and fluorescence intensity as a function of salt concentration for three different concentrations viz. 0.05 mg/ml, 0.06 mg/ml and 0.12 mg/ml. The data are taken from spectra shown in figure 4.

From figure 5a it has been observed that the FRET efficiency for 0.06 mg/ml and 0.12 mg/ml concentration are 48.2% and 13.5% respectively. If the FRET efficiency is observed to be higher than 48.2%, then the water will be soft water ( salt concentration < 0.06 mg/ml) whereas, if the efficiency lies in between 13.5% and 48.2% then the water will be moderately hard (0.06mg/ml < salt concentration < 0.12 mg/ml). On the other hand if the observed FRET efficiency is less than 13.5% then the water will be very hard (salt concentration > 0.12 mg/ml). Similarly, by observing, whether the fluorescence intensity of 579 nm peak (figure 5b) is greater than 392 or less than 232 it is possible to detect the soft water or very hard water. Again if the fluorescence intensity lies in between 232 and 392 then the water will be moderately hard water. Therefore with proper calibration it is possible to design a hard water sensor which can sense hard water very easily.

## 4. Conclusion

Fluorescence resonance energy transfer (FRET) between two fluorescent dyes Acriflavine and Rhodamine B were investigated successfully in solution in presence and absence of clay mineral particle laponite. UV-Vis absorption and fluorescence spectroscopy studies reveal that both the dyes present mainly as monomer in solution and there exist sufficient overlap between the fluorescence spectrum of Acf and absorption spectrum of RhB, which is a prerequisite for the FRET to occur from Acf to RhB. Energy transfer occurred from Acf to RhB in solution in presence and absence of laponite. The energy transfer efficiency increases in presence of clay laponite in solution. The maximum efficiency was found to be 78.17% for the mixed dye system (50% RhB + 50% Acf) in clay dispersion. In presence of $CaCl_2$ or $MgCl_2$ the FRET efficiency is decreased to 37.78% and 51.59% respectively. With suitable calibration of these results it is possible to design a hard water sensor that can sense the water hardness of the range 0.03 mg/ml-0.2 mg/ml.

**Acknowledgements**

The author SAH is grateful to DST, CSIR and DAE for financial support to carry out this research work through DST Fast-Track project Ref. No. SE/FTP/PS-54/2007, CSIR project Ref.

03(1146)/09/EMR-II and DAE Young Scientist Research Award (No. 2009/20/37/8/BRNS/3328).**References**

[1] H. Weingärtner, Water in Ullmann's Encyclopedia of Industrial Chemistry, Wiley–VCH, Weinheim, 2006 [December].

[2] T.J. Sorg, R.M. Schock, D.A. Lytle, Ion Exchange Softening: Effects on Metal Concentrations, J. Am. Water Works Assoc., 91 (1999) 85–97.

[3] P.B. Sweetser, C.E. Bncker, Spectrophotometric Titrations with Ethylenediaminetetraacetic Acid: Determination of Magnesium, Calcium, Zinc, Cadmium, Titanium, and Zirconium, Anal. Chem. 26 (1954) 195-199.

[4] H. Small, T.S. Stevens, W.C. Bauman, Novel ion exchange chromatographic method using conductimetric detection, Anal. Chem. 47 (1975) 1801-1809.

[5] P.R. Haddad, P.E. Jackson, Ion Chromatography—Principles and Applications, Elsevier, Amsterdam, 1991.

[6] M.D. Argiiello, J.S. Fntz, Ion-exchange separation and determination of calcium and magnesium, Anal. Chem. 49 (1977) 1595-1597.

[7] E. Gómez, J.M. Estela, V. Cerdà, Simultaneous spectrophotometric determination of calcium and magnesium in water, Anal. Chim. Acta 249 (1991) 513-518.15

[8] T.H. Förster, Experimentelle und theoretische Untersuchung des Zwischenmolekularen übergangs von Elektrinenanregungsenergie. Z. Naturforsch 4A (1949) 321-327.

[9] T.H. Förster, Transfer mechanisms of electronic excitation, Diss. Faraday Soc. 27 (1959) 7–71.

[10] G. Haran, Topical Review: Single-molecule fluorescence spectroscopy of biomolecular folding, J. Phys.: Condens. Matter, 15 (2003) R1291–R1317.

[11] R.B. Best, S.B. Fowler, J.L. Toca Herrera, J. Clarke, A simple method for probing the mechanical unfolding pathway of proteins in detail, Proc. Natl. Acad. Sci. U.S.A., 99 (2002) 12143–12148.

[12] B. Zagrovic, C.D. Snow, S. Khaliq, M.R. Shirts, V.S. Pande, Native-like Mean Structure in the Unfolded Ensemble of Small Proteins, J. Mol. Biol. 323 (2002) 153–164.

[13] R.J.H. Clark, R.E. Hester, Spectroscopy of Inorganic based Materials Advances in Spectroscopy, Wiley, New York, 1996.

[14] M.S. Csele, P. Engs, Fundamentals of Light and Lasers, Wiley, New York, 2004.

[15] P.D. Sahare, V.K. Sharma, D. Mohan, A.A. Rupasov, Energy transfer studies in binary dye solution mixtures: Acriflavine+Rhodamine 6G and Acriflavine+Rhodamine B, Spectrochim. Acta Part A 69 (2008) 1257 -1264.

[16] D. Dey, D. Bhattacharjee, S. Chakraborty, S.A. Hussain, Effect of nanoclay laponite and pH on the energy transfer between fluorescent dyes, J. Photochem. Photobiol. A-Chem., Accepted, 2012.

[17] S.A. Hussain, S. Chakraborty, D. Bhattacharjee, R.A. Schoonheydt, Fluorescence Resonance Energy Transfer between organic dyes adsorbed onto nano-clay and Langmuir–Blodgett (LB) films, Spectrochim. Acta Part A 75 (2010) 664-670.

[18] S.A. Hussain, R.A. Schoonheydt, Langmuir-Blodgett monolayers of cationic dyes in the presence and absence of clay mineral layers: N,N' -dioctadecyl thiacyanine, octadecyl rhodamine B and laponite, Langmuir 26 (2010) 11870-11877.

[19] T. Szabo, R. Mitea, H. Leeman, G.S. Premachandra, C.T. Johnston, M. Szekeres, I. Dekany, R.A. Schoonheydt, Adsorption of protamine and papine proteins on saponite, Clays Clay Miner. 56 (2008) 494-504.

[20] S.A. Hussain, S. Banik, S. Chakraborty, D. Bhattacharjee, Adsorption kinetics of a fluorescent dye in a long chain fatty acid matrix, Spectrochim. Acta Part A 79 (2011) 1642–1647.

[21] L.M. Felber, S.M. Cloutier, C. Kündig, T. Kishi, V. Brossard, P. Jichlinski, H.J. Leisinger, D. Deperthes, Evaluation of the CFP-substrate-YFP system for protease studies: advantages and limitations, Biotechniques 36 (2004) 878-885.

[22] R.M. Wachter, S.J. Remington, Sensitivity of the yellow variant of green fluorescent protein to halides and nitrate, Curr. Biol. 9 (1999) 628-629.

[23] S. Jayaraman, P. Haggie, R.M. Wachter, S.J. Remington, A.S. Verkman, Mechanism and Cellular Applications of a Green Fluorescent Protein-based Halide Sensor, J. Biol. Chem. 275 (2000) 6047-6050.

[24] K.Yoshioka, Y.Y. Yoshioka, F. Fleury, M. Takahashi, pH- and Salt-Dependent Self-Assembly of Human Rad51 Protein Analyzed as Fluorescence Resonance Energy Transfer between Labeled Proteins, J. Biochem. 133 (2003) 593–597.

[25] D. Seth, D. Chakrabarty, A. Chakraborty, N. Sarkar, Study of energy transfer from 7-amino coumarin donors to rhodamine 6G acceptor in non-aqueous reverse micelles, Chem. Phys. Lett. 401 (2005) 546-552.

[26] S.A. Hussain, S. Chakraborty, D. Bhattacharjee, R.A. Schoonheydt, Fluorescence Resonance Energy Transfer between organic dyes adsorbed onto nano-clay and Langmuir–Blodgett (LB) films, Spectrochim. Acta Part A 75 (2010) 664-670.

[27] R.A. Schoonheydt, Smectite-Type clay minerals as nanomaterials, Clay Clay Miner. 50 (2002) 411–420.

[28] R.H.A. Ras, B. van Duffel, M. Van der Auweraer, F.C. De Schryver, R.A. Schoonheydt, in: Proceedings of the 12 th International Clay Conference, Bahia Blanca, Argentina, 2001.

[29] C.E. Marshall, Multifunctional ionization as illustrated by the clay minerals, Clays and Clay Minerals 327(1954) 364-385.

**Table 1**

Values of energy transfer efficiency (E %) for Acf and RhB mixture (1:1 volume ratio) in different conditions. The salt concentration was 0.06 mg/ml (moderately hard water). Dye concentration was $10^{-6}$M and clay concentration was 2 ppm.

| Samples | E% |
|---|---|
| Acf+RhB | 11.37 |
| Acf+RhB+ $CaCl_2$ | 1.7 |
| Acf+RhB+ $MgCl_2$ | 5.2 |
| Acf+RhB+ $MgCl_2$ +$CaCl_2$ | 4.38 |
| Acf+RhB+clay | 78.17 |
| Acf+RhB+ $CaCl_2$ with clay | 37.78 |
| Acf+RhB+ $MgCl_2$ with clay | 51.59 |
| Acf+RhB+ $MgCl_2$ +$CaCl_2$ with clay | 48.18 |

**Table 2**

Values of energy transfer efficiency (E %) for Acf and RhB mixture (1:1 volume ratio) at different salt concentration in presence of clay. Dye concentration was $10^{-6}$M and clay concentration was 2 ppm.

| Salt concentration in mg/ml ($MgCl_2$ +$CaCl_2$) | E% |
|---|---|
| 0.03 | 73.73 |
| 0.05 | 68.34 |
| 0.06 | 48.18 |
| 0.08 | 21.57 |
| 0.12 | 13.48 |
| 0.20 | 07.38 |

**Figure captions**

**Fig. 1.** Fluorescence spectra of Acf+RhB (1:1 volume ratio) in water solution (1), with $MgCl_2$ (2), $CaCl_2$ (3), and $CaCl_2$+$MgCl_2$ (4), pure Acf (5), pure RhB (6). Dye concentration was $10^{-6}$M and salt concentration was 0.06 mg/ml.

**Fig. 2.** Fluorescence spectra of Acf+RhB (1:1 volume ratio) in clay suspension (1), with $MgCl_2$ (2), $CaCl_2$ (3), and $CaCl_2$+$MgCl_2$ (4) pure Acf with clay (5), pure RhB with clay (6). Dye concentration was $10^{-6}$M and clay concentration was 2 ppm and salt concentration was 0.06 mg/ml.

**Fig. 3.** Schematic representation of FRET between Acf and RhB in presence of clay and salt.

**Fig. 4.** Fluorescence spectra of Acf+RhB (1:1 volume ratio) in clay dispersion with $CaCl_2$ + $MgCl_2$ of concentration 0.06 mg/ml (3), 0.05 mg/ml (2) and 0.12 mg/ml (1). Inset shows the variation in FRET efficiency with the increasing concentration of $CaCl_2$ + $MgCl_2$ from 0.03 mg/ml to 0.20 mg/ml. Dye concentration was $10^{-6}$M and clay concentration was 2 ppm.

**Fig. 5a.** FRET efficiency of Acf and RhB mixture for the different concentration of $CaCl_2$ + $MgCl_2$ in presence of clay. (Values of FRET efficiencies were calculated from the spectra of figure 4).

**Fig. 5b.** Fluorescence intensity of 579 nm peak for the different concentration of $CaCl_2$ + $MgCl_2$ in presence of clay. (Values of FRET efficiencies were calculated from the spectra of figure 4).

**Table captions**

**Table1**

Values of energy transfer efficiency (E %) for Acf and RhB mixture (1:1 volume ratio) in different conditions. The salt concentration was 0.06 mg/ml (moderately hard water). Dye concentration was $10^{-6}$M and clay concentration was 2 ppm.

**Table 2**

Values of energy transfer efficiency (E %) for Acf and RhB mixture (1:1 volume ratio) at different salt concentration in presence of clay. Dye concentration was $10^{-6}$M and clay concentration was 2 ppm.

**Figures:**

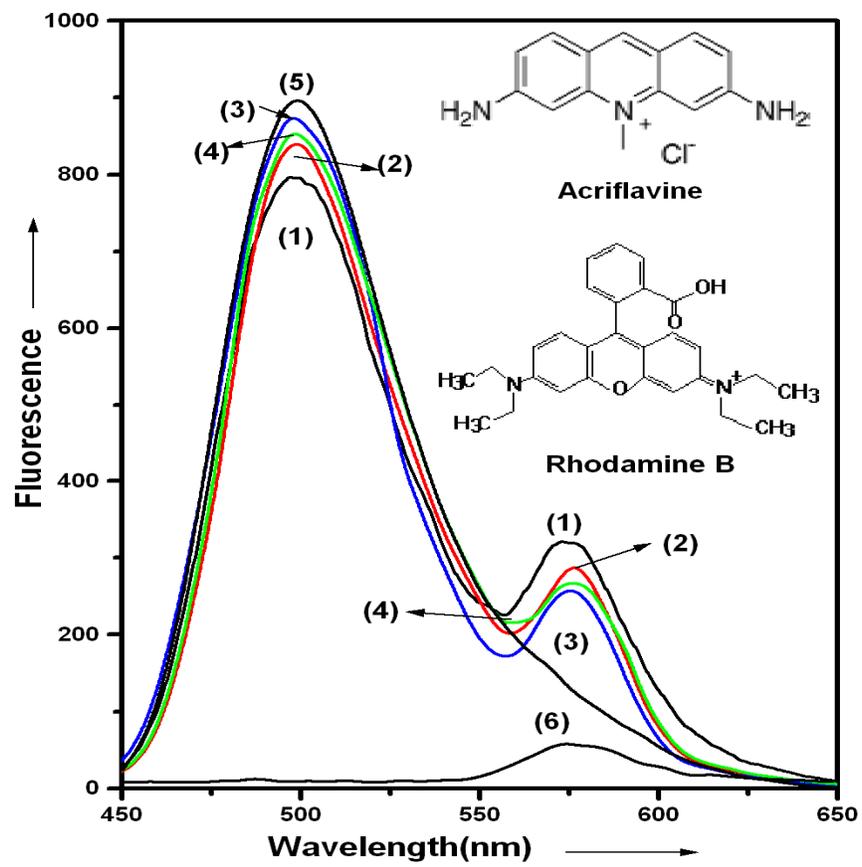

**Fig. 1.**

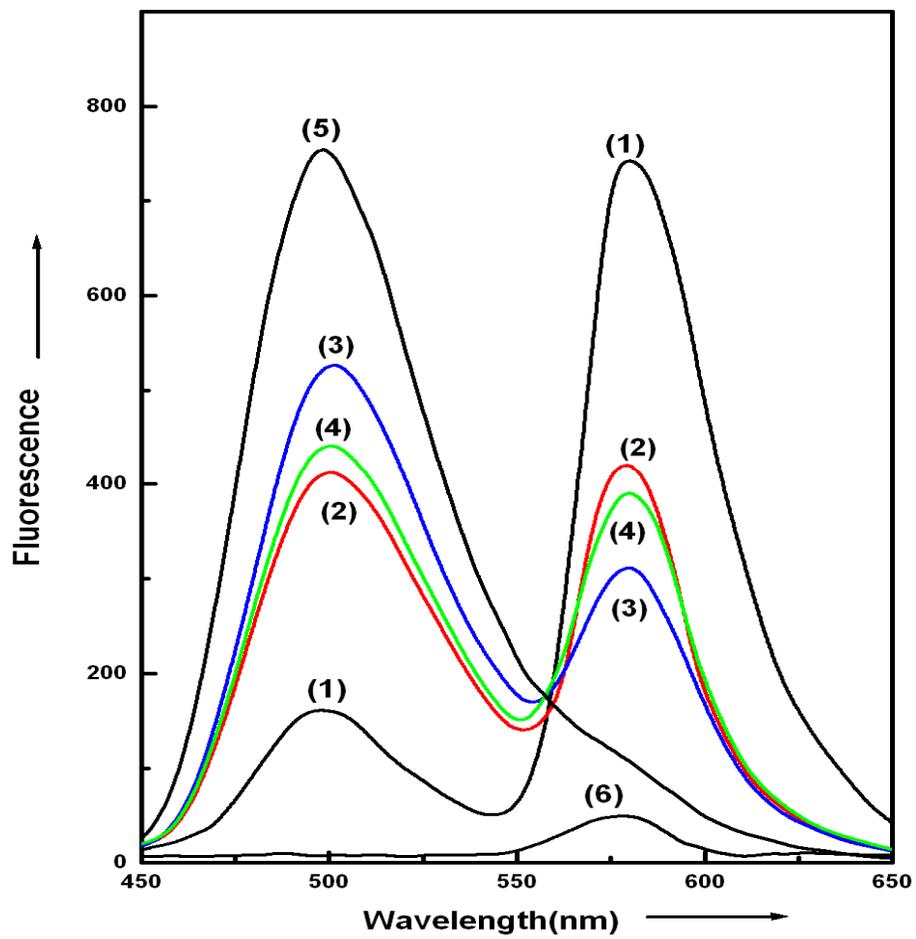

**Fig. 2.**

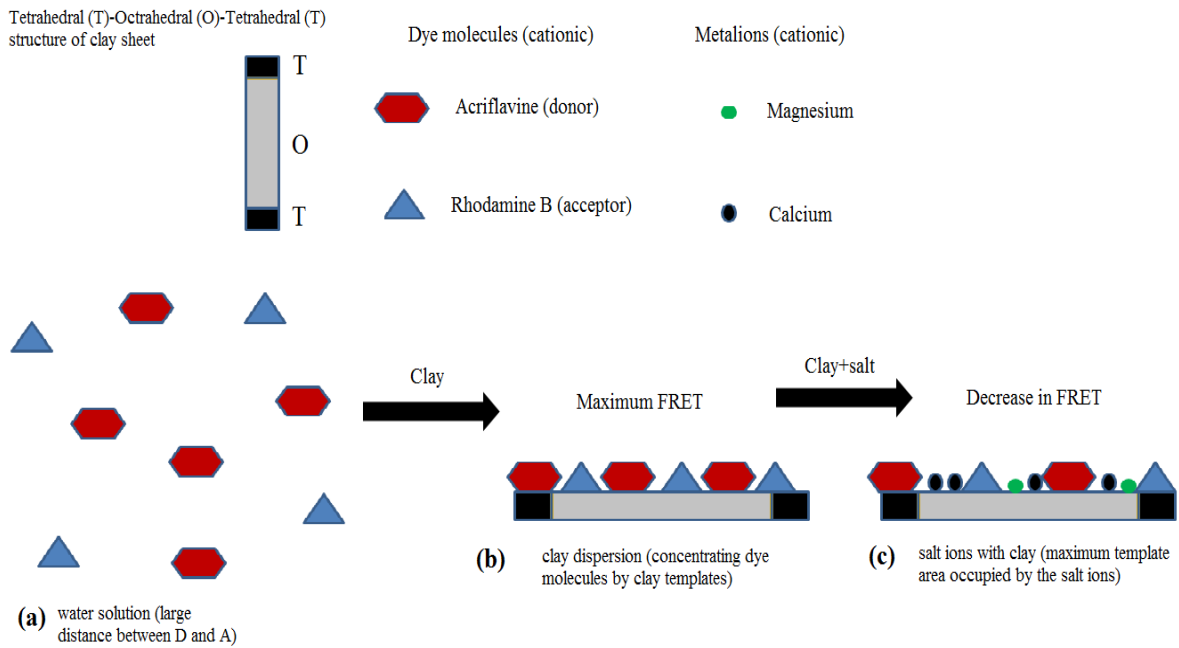

**Fig. 3.**

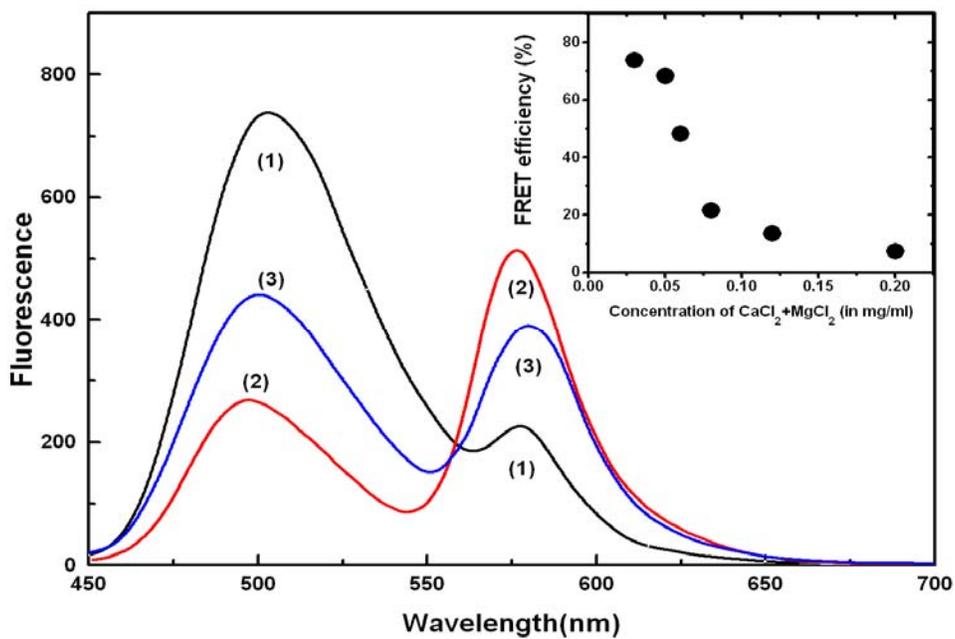

**Fig. 4.**

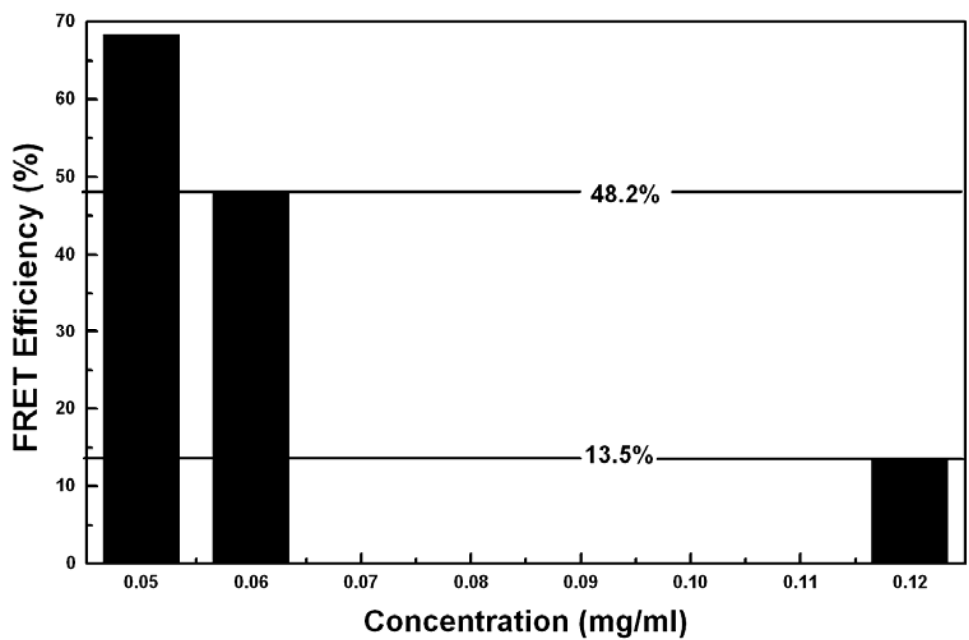

**Fig. 5a.**

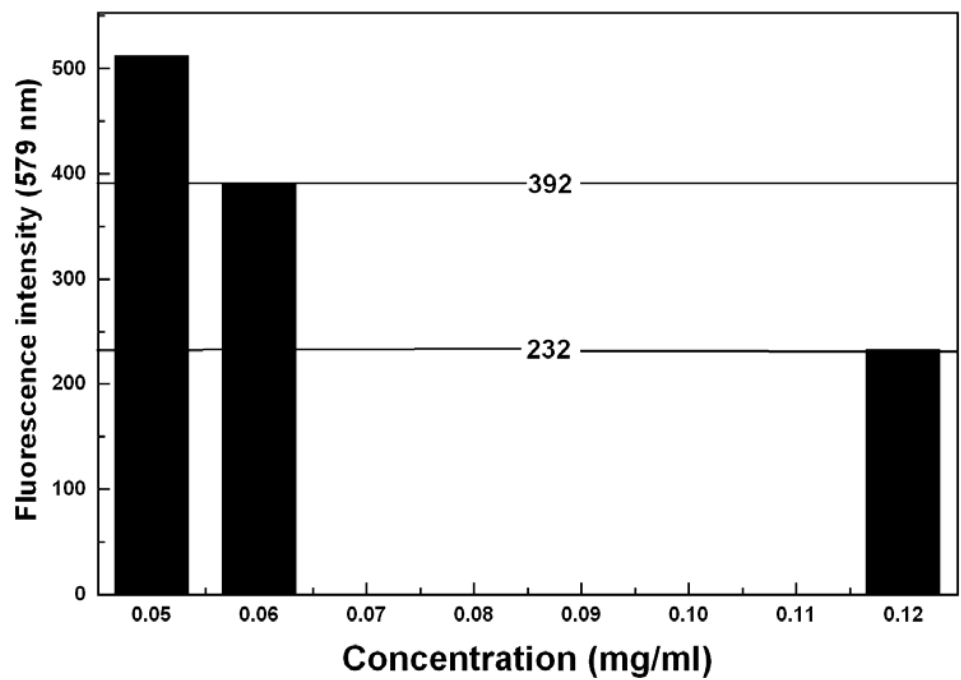

**Fig. 5b.**

**Authors Biography:**

**Prof. D. Bhattacharjee** (M.Sc, Kalyani University & Ph.D, IACS, India) is a Professor in the department of Physics, Tripura University,India. His major fields of interest are preparation and characterization of ultra thin films by Langmuir-Blodgett & Self-assembled techniques. He visited Finland and Belgium for postdoctoral research. Ha has undertaken several research projects. He has published more than 50 research papers in different national and international journals and attended several scientific conferences in India and abroad.

**Dr. S. A. Hussain** (M.Sc. 2001 & Ph.D, 2007, Tripura University, India) is an Assistant Professor in the Department of Physics, Tripura University. His major fields of interest are preparation and characterization of ultra thin films by Langmuir-Blodgett & Self-assembled techniques. He visited Postdoctoral Fellow of K.U. Leuven, Belgium (2007-08). He received Jagdish Chandra Bose Award 2008-2009, TSCST, Govt. of Tripura; Young Scientist Research Award by DAE, Govt of India and CSIR, Govt. of India. He has published 41 research papers in different national and international journals and attended several scientific conferences in India and abroad.

**Mr. Dibyendu Dey** (M.Sc 2009, Tripura University, India) is working as a Research scholar in department of physics, Tripura University. His major fields of interest are preparation and characterization of ultra thin films by Langmuir-Blodgett & Self-assembled techniques. He has published 3 research papers in various international journals and attended several scientific conferences in India.

**Mr. Sekhar Chakraborty** (M.Sc 2006, Tripura University, India) is working as a Senior Research Fellow in Department of Physics, Tripura University. His major fields of interest are preparation and characterization of organo-clay hybrid ultra thin films by Langmuir-Blodgett & Self-assembled techniques. He has already published 8 research papers in reputed international journals and attended several national and international scientific conferences in India.